\documentstyle[prl,aps,multicol,epsf]{revtex}

\newcommand{\br}{{\bf r}}
\newcommand{\bq}{{\bf q}}
\newcommand{\hL}{\widehat{\cal L}}
\begin{document}
%\draft 
\title{Thermodynamics of Mesoscopic Vortex Systems in 1+1 Dimensions} 
\author{Chen Zeng$^{(1)}$, P.L. Leath$^{(1)}$, and Terence Hwa$^{(2),}$\cite{ucsd}}
\address{$^{(1)}$Department of Physics and Astronomy, Rutgers University, 
Piscataway, NJ 08854\\
$^{(2)}$ Center for Study in Physics and Biology, Rockefeller University, 
New York, NY 10021}

\date{Received: \today}

\maketitle  

\widetext
\begin{abstract} 
The thermodynamics of a disordered planar vortex array is studied
numerically using a new polynomial algorithm which circumvents
slow glassy dynamics. Close to the glass transition, 
the anomalous vortex displacement is found to agree well with 
the prediction of the renormalization-group theory. Interesting behaviors
such as the universal statistics of magnetic susceptibility variations
are observed in both the dense and dilute regimes of this mesoscopic
vortex system.

\end{abstract} 

%\date\today

\begin{multicols}{2}
\narrowtext

%%%%%%%%%%%%%%%%%%%%%%%%%%%%%%%%%%%%%%%%%%%%%%%%%%%%%%%%%%%%%%%%%%%%%%%%%%%%%%
%\paragraph*{Introduction}
%%%%%%%%%%%%%%%%%%%%%%%%%%%%%%%%%%%%%%%%%%%%%%%%%%%%%%%%%%%%%%%%%%%%%%%%%%%%%%

The behavior of vortices in dirty type-II superconductors has been a subject
of intense studies in the last decade~\cite{blatter}.
Aside from the obvious technological significance of vortex pinning,
understanding the physics of such interacting many-body systems in the presence
of quenched disorder is a central theme of modern condensed matter physics.
Similarities between the randomly-pinned vortex system and the more familiar
mesoscopic  electronic systems~\cite{meso} are highlighted by 
a recent experimental study of a planar vortex array threaded through
a thin crystal of $2$H-NbSe$_2$  by Bolle {\it et al.}~\cite{bolle}. 
Interesting behaviors, including the sample-dependent magnetic responses 
known as ``finger prints'', have been observed for such a mesoscopic 
vortex system.

The disordered planar vortex array is well studied
theoretically~\cite{rough,mpaf,natt,gld,hnv,usf}. 
It is one of the few disorder-dominated
systems for which quantitative predictions can be made, including 
a finite-temperature ``vortex glass'' phase~\cite{mpaf} characterized by
anomalous vortex displacements~\cite{rough}, and universal variation 
of magnetic susceptibility~\cite{usf}.
However, until the work of Bolle {\it et al.}, there were hardly any 
experimental studies of this system, with difficulties stemming
partly from the weak magnetic signals in such 2d systems. 
Also, numerical simulations have been
limited by the slow glassy dynamics~\cite{2d_finite}, although the availability
of special optimization algorithms did lead to the elucidation of the zero-temperature
problem in recent years~\cite{2d_zero}.
In this letter, we describe numerical studies of the
thermodynamics of the vortex glass via a mapping
to a discrete dimer model with quenched disorder.
A new polynomial algorithm for the dimer problem
circumvents the glassy dynamics and enables us to study
large systems at finite temperatures. Our results obtained in the dilute 
(single-flux-pinning)
regime compare well with the  experiment by Bolle {\it et al.}~\cite{bolle},
while those obtained in the collective-pinning regime 
strongly support the renormalization-group theory of the 
vortex glass, including its prediction
of universal susceptibility  variation~\cite{usf}.

%%%%%%%%%%%%%%%%%%%%%%%%%%%%%%%%%%%%%%%%%%%%%%%%%%%%%%%%%%%%%%%%%%%%%%%%%%%%%%
\paragraph*{The Model:}
%%%%%%%%%%%%%%%%%%%%%%%%%%%%%%%%%%%%%%%%%%%%%%%%%%%%%%%%%%%%%%%%%%%%%%%%%%%%%%
The dimer model consists of all complete dimer coverings $\{D\}$ on a
square lattice $\cal L$ as illustrated in Fig.~1(a). The partition
function is   
\begin{equation} 
Z=\sum_{\{D\}} \exp\left[-\sum_{<ij>\in D} \epsilon_{ij}/T_d\right]
\;\; , 
\label{eq1}
\end{equation} 
where the sum in the exponential 
is over all dimers of a given covering, and $T_d$ is the dimer temperature.
Quenched disorder is introduced
via random bond energies $\epsilon_{ij}$, chosen independently and
uniformly in the interval $(-\frac12,\frac12)$. 

\begin{figure}
  \begin{center}
    \leavevmode
    \epsfxsize=8cm
    \epsffile{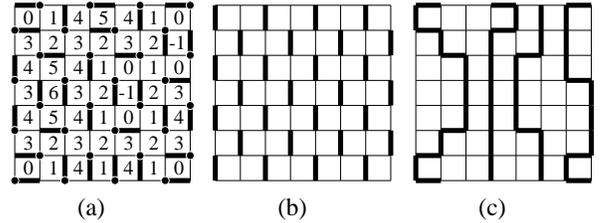}
  \end{center}
\caption{
(a) Snapshot of a dimer covering (thick bonds) together 
with the associated ``height'' values $h(\br)$
 for a lattice of size $L=8$ at temperature $T_d=1.0$. 
(b) Dimer covering of the fixed reference. (c) Vortex line configuration
(thick lines) obtained as the difference between (a) and (b).    
}
\label{fig_1}
\end{figure}

The dimer model is related to the planar vortex-line array via
the well-known mapping to the solid-on-solid (SOS) model (see Figs.~1):
Take the centers of the square cells of ${\cal L}$ to form the dual square 
lattice $\hL$. Orient all bonds of $\hL$ such
that the elementary squares of $\hL$ that enclose the sites of the
chosen sublattice of ${\cal L}$ (indicated by the solid dots in Fig.~1(a)) 
are circled counterclockwise. It is now possible to assign a single-valued
``height'' function $h(\br)$ on the lattice points $\br$ of $\hL$, such that
the difference of every pair of neighboring heights
across the oriented bonds is $-3$ if a dimer is crossed and $+1$ otherwise. 
For the dimer covering of Fig.~1(a), the values of the associated
height function are shown at their respective positions.
In terms of the height configuration $h(\br)$, the partition function 
(\ref{eq1}) can be written alternatively
as $Z = \sum_{\{h(\br)\}}
e^{-\beta {\cal H}[h]}$, where the SOS Hamiltonian
takes the following form in the continuum limit,
\begin{equation} 
\beta {\cal H}=\int d^2 \br\left[
\frac{K}{2} (\nabla h)^2
-{\bf f}({\bf r})\cdot\nabla h
+g\cos\left(G h \right)
\right]
\;\; .
\label{eq2}
\end{equation} 
Here, $K$ is an effective stiffness caused by the
inability of a tilted surface to take as much advantage of the 
low weight bonds as a flatter surface, and 
${\bf f}({\bf r})$ is a random local tilt bias. 
The periodicity of the cosine potential in (\ref{eq2}) is given by $G=2\pi/4$
since the smallest ``step'' of this height profile is four.
In the present context of a randomly pinned vortex array,
$h(\br)$ describes the coarse-grained 
displacement field of the vortex array with respect to 
a uniform reference state at the same vortex line density; see Fig.~1(c)
and Refs.~\cite{mpaf,hnv}.
This mapping between the dimer model and the planar vortex array
allows us to compute statistical properties of the vortex array by 
monitoring appropriate quantities of the dimer model. The latter can be
accomplished using a polynomial algorithm described below.

%%%%%%%%%%%%%%%%%%%%%%%%%%%%%%%%%%%%%%%%%%%%%%%%%%%%%%%%%%%%%%%%%%%%%%%%%%%%%%
\paragraph*{The Algorithm:}
%%%%%%%%%%%%%%%%%%%%%%%%%%%%%%%%%%%%%%%%%%%%%%%%%%%%%%%%%%%%%%%%%%%%%%%%%%%%%%
Computing the partition function of complete dimer coverings
on an {\em arbitrary} weighted lattice is likely to be algorithmically
intractable\cite{dimer_NP}. Weights here refer to the Boltzmann factors
$w_{ij} \equiv \exp(-\epsilon_{ij}/T_d)$ on the bonds. A weighted {\it planar} 
lattice $\cal G$ can however be {\em oriented}, denote by  $\vec{\cal G}$,
so that the 
square root of the determinant of the weighted adjacency matrix defined 
on $\vec{\cal G}$ yields the partition function (\ref{eq1})
{\em exactly}~\cite{dimer_pfaff}. Computation of the determinant can be 
achieved by various row or column reduction schemes
in time that grows {\em polynomially} with the matrix size.

Recently, Propp and coworkers\cite{dimer_shuffle} furnished
the above algebraic reduction with an elegant graphical 
interpretation for the case of bipartite planar 
lattices~\cite{dimer_aztec}. The partition 
function $Z_L$ on a lattice of linear size $L$ is related to $Z_{L-1}$
as $Z_L(\{w\})=C(\{w\}) Z_{L-1}(\{w^\prime\})$ after a simple weight 
transformation $\{w\}\rightarrow\{w^\prime\}$. The prefactor
$C(\{w\})$ is independent of dimer coverings. 
The partition function is obtained in a ``deflation'' process in which
the above recursive procedure is carried out down to $L=0$ with
$Z_0=1$. This deflation process can also be reversed in an
``inflation'' process where a dimer covering at size $L-1$ can be used
to {\it stochastically} generate a dimer covering at size $L$ according to 
$Z_L$ already obtained. Repeating the inflation process thus generates
{\it uncorrelated} ``importance samplings'' of the dimer configurations, or
equivalently the equilibrium height configurations, without the need to 
run the slow relaxational dynamics.
The ensuing numerical results are obtained by taking various measurements
of the height configurations generated this way.
A somewhat inconvenient feature of this approach is that the algorithm
requires open boundary condition on the dimer model; this in turn 
fixes the total number of vortex lines, e.g.,  to $L/2$ 
on a $L\times L$ lattice (see Fig.~1).

\paragraph*{Numerical Results:} 
Since the temperature of the vortex array, given by $K^{-1}$ in (\ref{eq2}), 
is generally different from the dimer temperature $T_d$, 
we first need to calibrate the temperature scale.
To do so, we exploit a statistical rotational symmetry~\cite{symmetry,usf} of the 
system (\ref{eq2}) which guarantees that at large scales, the effective $K$ 
is the same as that of the pure system. In particular, $K$ can be obtained
by measuring the disorder-averaged thermal fluctuation of the displacement 
field $\overline{W_T^2}\equiv \overline{
\langle\prec h_T^2({\bf r})\succ - \prec h_T({\bf r})\succ^2\rangle}
= (2\pi K)^{-1} \ln(L)$, where $h_T({\bf r})
\equiv h({\bf r}) - \langle h({\bf r})\rangle$ measures the thermal distortion
superposed on the distorted ``background'' 
$h_D({\bf r}) \equiv 
\langle h({\bf r}) \rangle$ by the 
quenched disorder. We used 
$\prec{...}\succ$,
$\langle{...}\rangle$, and
overline to denote spatial, thermal, and disorder averages
respectively. 
With our polynomial algorithm, we were able to perform 
thorough disorder averages for equilibrated systems of sizes 
up to $512 \times 512$. To reduce boundary effects, 
we focus on the central $L/2\times L/2$ piece of the system
and compute its displacement fluctuations.
Fig.~2(a) illustrates the dependence of $\overline{W_T^2}(L)$ for
various dimer temperatures
$T_d$. The linear dependence on $\ln(L)$ is apparent. Identifying the
proportionality constant with $(2\pi K)^{-1}$, we obtain the empirical
relation between $K^{-1}$ and $T_d$ shown in Fig.~2(b). 
Note that our result 
recovers the exact relation $K^{-1}(T_d\to\infty) = 16/\pi$ for  
the dimer model without disorder~\cite{exact}.
Since the glass transition of the system (\ref{eq2}) is expected to
occur at temperature
$K_g^{-1} =4\pi/G^2 = 16/\pi$, our system is glassy for the
entire range of dimer temperatures.

\begin{figure}
  \begin{center}
    \leavevmode
    \epsfysize = 1.5in
    \epsffile{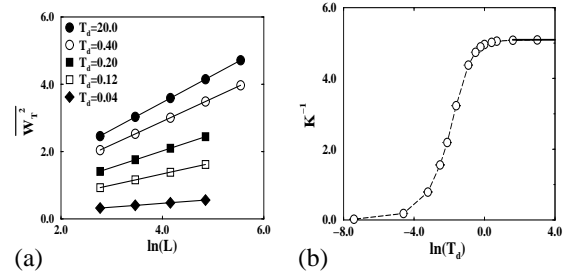}
  \end{center}
\caption{(a) The disorder-averaged thermal displacement fluctuation 
$\overline{W_T^2}(L)$ 
for various dimer temperatures $T_d$. Each data point was obtained by 
generating $10^3$ thermal samplings for each of $10^3$ 
disorder realizations. The straight lines are fit to 
$\overline{W_T^2} = (2\pi K)^{-1} \ln(L) + {\rm const}$.
 (b) The extracted relation between  $K^{-1}$ and $T_d$. 
The horizontal line indicates the asymptotic value of $K^{-1}$
as $T_d \to\infty$.
}
\label{fig_2}
\end{figure}

%\paragraph{Displacement fluctuations} 

A striking feature of the vortex glass phase is the anomalous fluctuation
of the disorder-induced displacement,  
$\overline{W_D^2}\equiv \overline{
\langle\prec h_D^2({\bf r})\succ - \prec h_D({\bf r})\succ^2\rangle}$.
The renormalization group (RG) theory~\cite{rough} predicts
that $\overline{W_D^2}(L) \simeq C_2 \ln^2(L)$ for large $L$, 
with the proportionality constant $C_2$ depending quadratically on the reduced temperature 
$\tau \equiv K_g/K -1$ just below the glass-transition temperature $K_g^{-1}$. 
We show $\overline{W_D^2}(L)$ vs.~$\ln(L)$ for different temperatures in Fig.~3(a). 
The forms of $\overline{W_D^2}(L)$ are well described by 
a quadratic function of $\ln(L)$ as predicted.
The solid lines are least-square fit to the form
$\overline{W_D^2}(L) = C_2 \ln^2(L) + C_1 \ln(L) + C_0$.
\begin{figure}[t]
  \begin{center}
    \leavevmode
    \epsfysize=1.5in
    \epsffile{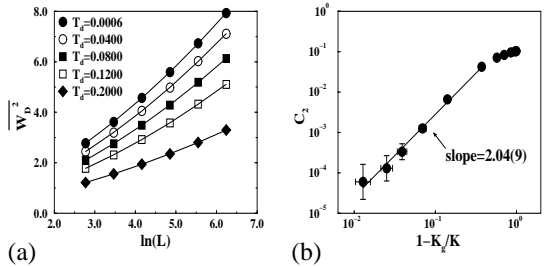}
  \end{center}
\caption{(a) The anomalous displacement fluctuation $\overline{W_D^2}$ 
at various dimer temperatures averaged over $10^3$ disorder realizations. 
The lines are fits to the quadratic dependence on $\ln(L)$ expected from the
RG theory. 
(b) Log-log plot of the quadratic coefficient $C_2$ (extracted from (a)) 
vs.~the reduced temperature $\tau = K_g/K - 1$.
The best power law fit yields an exponent of $2.04(9)$. 
}
\label{fig_3}
\end{figure}

To probe the RG prediction further, we examine the temperature dependence
of the quadratic coefficient $C_2$. Plotting $C_2$ extracted from the quadratic fit
in Fig.~3(a) vs. the reduced temperature $\tau = K_g/K - 1$ using 
the vortex temperature $K^{-1}$ obtained in Fig.~2(b) and 
the exact glass-transition temperature $K_g^{-1}=16/\pi$,  
we obtain the data shown in Fig.~3(b). A quadratic dependence of the coefficient
$C_2$ on the reduced temperature $\tau$ is clearly demonstrated 
by the data, thereby  
providing strong support for the RG theory of the vortex
glass.

%\paragraph{Universal susceptibility fluctuations} 
Another interesting feature of the vortex glass phase
is the sample-to-sample variation of the  magnetic 
susceptibility $\chi$. It was predicted~\cite{usf} that the fractional variance,
\begin{equation}
{\rm var}[\chi]/\overline{\chi}^2 \stackrel{L\to\infty}{\rightarrow} D |\tau|
\qquad {\rm for} \quad 0 < -\tau \ll 1,
\label{Fsus}
\end{equation} 
is {\em universal},
with $D$ being a computable, size-independent constant of order unity.
Due to the constraint of fixed vortex density when using the dimer representation,
it is not easy to probe the magnetic susceptibility directly
by varying vortex density.
Instead, we use the fluctuation-dissipation relation,
e.g.,  $\chi \propto \langle (\partial_x h)^2 \rangle
- \langle \partial_x h \rangle^2$. To circumvent lattice effects, we take as definition
$\chi(L) \equiv \bq^2 
(\langle \hat{h}_{\bq} \hat{h}_{-\bq} \rangle 
-\langle \hat{h}_{\bq}\rangle \langle \hat{h}_{-\bq}\rangle)$, where $\hat{h}_{\bq}$ 
is the Fourier transform of $h(\br)$, and the right-hand side is evaluated 
at $(q_x,q_y)=(1/L,0)$. 
As shown in Fig.~4, the fractional variance is indeed size-independent over
the range $L=32$ to $L= 256$ examined. Its temperature dependence can again be
deduced from the relation between $K^{-1}$ and $T_d$ given in Fig.~2(b); it is
well described by the linear form (\ref{Fsus})
close to the glass transition with $D= 0.454(5)$. 

%\paragraph{Single fluxline pinning regime} 

The universal variation of magnetic susceptibility  is reminiscent of
the phenomena of universal conductance fluctuation much studied
in mesoscopic electronic systems. 
This mesoscopic vortex system in fact exhibits interesting fluctuations
even deep into the dilute regime, i.e., close to the threshold field $H_{c1}$.
But to study the
\begin{figure}[t]
  \begin{center}
    \leavevmode
    \epsfysize= 1.5in
    \epsffile{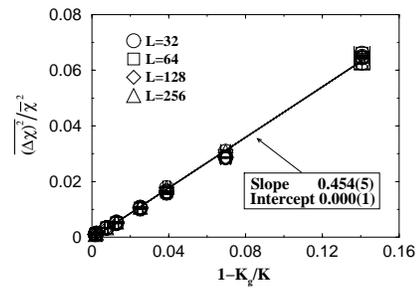}
  \end{center}
\caption{Fractional variance as a function of the reduced temperature,
$\tau$, for $L=32, 64, 128$, and $256$ averaged over $10^3$ disorder
realizations.
The solid line is a least-square linear fit for $|\tau| < 0.15$.
}
\label{fig_4}
\end{figure}
\noindent vortex behavior there, we must
relax the constraint of fixed vortex number. To do so, 
we focus on  a $L_x\times L_y$   
located at the center of the $L\times L$ lattice.
Random bond energies are assigned only to this ``inner'' region.
While the total number of vortex lines must be fixed, 
the number of lines  in the inner region can be changed by adding
an extra bond energy $\delta\epsilon_{ij}=
(-1)^{i+j} v $ in a staggered fashion to all vertical bonds 
of the staggered reference dimer pattern (Fig.~1(b))
in this region.  This has the effect of increasing the ``self-energy''
of the vortex lines such that  when  $v$ is sufficiently 
large, i.e., for $v \ge v_c$, all vortex lines will be 
{\em expelled} from the inner region to the outer region; see Fig.~5.
A {\em dilute} regime in the inner region is then achieved for small $\delta v
\equiv v_c - v$. This setup could be interpreted as embedding the
dirty type-II  superconductor of focus within another pure type-II superconductor 
which has a much smaller $H_{c1}$. In the limit $L \gg L_x$, the 
outer region can be regarded as an infinite reservoir in which 
the vortex density change is negligible. The response we measure
will thus come essentially from the inner region.

\begin{figure}
  \begin{center}
    \leavevmode
    \epsfxsize=3.4in
    \epsffile{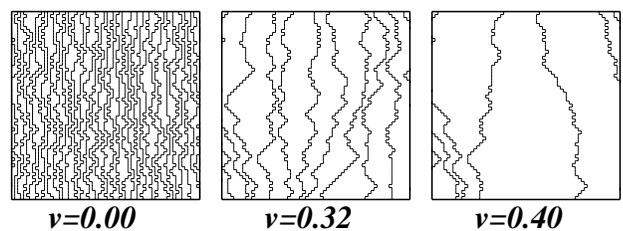}
  \end{center}
%\vspace{1.5in}
\caption{Snapshots of vortex line configurations in the inner region
at different repulsion 
strength $v$ for a given disorder realization and $T_d=0.01$. Quenched 
disorder exists only in the inner region of size $L_x=L_y=L/2 =64$. 
}
\label{fig_5}
\end{figure}

In Fig.~6(a), 
we plot the thermal averaged number of vortex lines $\langle n \rangle$
in a thin strip of
size $L_x = 16$, $L_y = 256$ embedded in a $512\times 512$ square lattice.
We vary the repulsion strength  $v$, 
holding the temperature fixed at $T_d = 0.01$ (corresponding to only
$\sim 10\%$ of the glass  transition temperature).
The solid squares are data
taken from a {\em given} realization of disorder. We see  
a disordered stair-case structure in the dependence of $\langle n \rangle$ on
$\delta v$, similar to experimental magnetic response curves 
observed by Bolle {\it et al}~\cite{bolle}.
The step width dispersion results from a delicate energetic balance between
the line-line repulsion and the random pinning; it can 
be regarded as a {\em finger print} of this disordered sample.
By averaging such stair-case-response curves over $100$ realizations. 
we obtain results indicated by the open squares in Fig.~6(a).
We can extract a linear dependence, i.e., $\overline{\langle n \rangle}
\sim \delta v$ (except for finite-size rounding very close to the threshold).
This finding agrees well with the experimental 
result of Bolle {\it et al}\cite{bolle}. It should be contrasted with 
the very different behavior
when the disorder is absent. The latter is shown in Fig.~6(b) for a system
which is twice larger in all linear dimensions. The data are well described
by a square-root singularity,  $\langle n \rangle \sim (\delta v)^{1/2}$,
as expected from the theory of Pokrovsky and Talapov~\cite{pt}.
The origin of the anomalous linear dependence of 
$\overline{ \langle n \rangle }$ on $\delta v$ in the random case
has been discussed previously~\cite{kardar,hnv};  
it can be readily understood from the properties of 
a single vortex line pinned in a two-dimensional random potential.

\begin{figure}
  \begin{center}
    \leavevmode
    \epsfxsize=7.0cm
    \epsffile{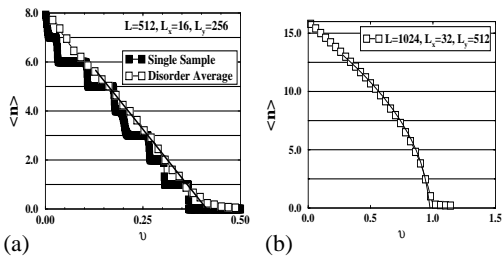}
  \end{center}
\caption{
(a) Average number of vortex lines $\langle n \rangle$ vs. different
repulsion strength $v$ at temperature $T_d=0.01$ 
in a narrow $L_x\times L_y$ stripe. 
The solid squares are for a given disorder realization and the 
open squares denote results from average over $100$ realizations.
The line is a fit to linear dependence in the vicinity of the threshold 
$v_c$.
Slight rounding near $v_c$ is due to the small but still finite 
ratio $L_x/L$. 
(b) $\langle n \rangle$ vs. $v$ in the absence of disorder.
The line
is a fit to the square-root form expected in the vicinity of $v_c$.
}
\label{fig_6}
\end{figure}

%%%%%%%%%%%%%%%%%%%%%%%%%%%%%%%%%%%%%%%%%%%%%%%%%%%%%%%%%%%%%%%%%%%%%%%%%%%%%%
\paragraph*{Conclusion.}
%%%%%%%%%%%%%%%%%%%%%%%%%%%%%%%%%%%%%%%%%%%%%%%%%%%%%%%%%%%%%%%%%%%%%%%%%%%%%%
Finite temperature simulations were performed on 
a disordered dimer model to study various equilibrium properties of a planar 
array of vortex lines. Numerical results on the anomalous displacement
fluctuation strongly support  the renormalization-group theory of 
vortex glass. 
Universal susceptibility variations in the collective pinning 
regime were observed in accordance with the theory; 
critical behaviors near the glass transition 
compared well with analytic predictions. Suitable 
modification of the dimer model allowed a direct
study of the dilute regime where a few vortex lines penetrate the system.
Disordered stair-case-like magnetic finger prints were obtained; the
statistics are in agreement with 
theoretical predictions, and drastically different from the well-known
behavior of the pure system. The numerical findings in the dilute regime
are also in agreement with recent experiment
on  a micron-sized single crystal of $2$H-NbSe$_2$.
It is hoped that the present study will stimulate further experimental
investigations of the fascinating physics of mesoscopic vortex systems.

TH acknowledges the financial support of the ONR Young Investigator
Program and the NSF through Grant No.~DMR-9801921.
CZ also acknowledges useful discussions with
D.S.~Fisher and D.A.~Huse. 

%\eject

\end{multicols}

\begin{thebibliography}{99}
\bibitem{dimer_aztec}
The graphical scheme was described for a square lattice of Aztec
diamond shape\cite{dimer_shuffle}. This poses no real restriction however 
since any bipartite planar graph can always be embedded in an Aztec diamond 
of sufficient size in which the dimer covering outside the graph is
frozen and makes no contribution to the partition function.
\end{thebibliography}
\end{document}